# Can the gravitational waves emitted by Eta Carinae be detected?

Odylio D. Aguiar [1] *

[1] Instituto Nacional de Pesquisas Espaciais, 12227-010 São José dos Campos, São Paulo, Brazil
* odylio.aguiar@inpe.br; Tel.: +55-12-98113-3733

**Abstract**
In this article, we propose a search for gravitational waves emitted by Eta Carinae using the PTA technique—not to detect a background, but rather a strong, isolated source (a task not yet accomplished). We suggest including pulsars PSR J09520607, PSR J17191438, PSR J23222650, and PSR J1909-3744 in the search network, as they exhibit relatively high residual-to-pulse-width ratios and spatial positions that favor the detection of tensor polarizations. The high eccentricity (0.9) of the Eta Carinae binary orbit may facilitate signal identification within the data due to the presence of harmonics in the emitted gravitational waves. It is possible that some of the raw data have already been recorded and are available; the remaining step is to analyze them against a model of the gravitational wave signal arriving at Earth from Eta Carinae.

## 1. Introduction

Eta Carinae (η Carinae, abbreviated as η Car) is one of the best-studied stellar systems [1, 2, 3, 4]. It contains at least two stars located about 7,500 light years (2,300 parsecs) away from Earth in the constellation Carina. It is a system that is constantly changing. In 1837, its brightness increased, indicating some type of eruption. It was once the brightest star in the night sky in March 1843, before disappearing to the naked eye 13 years later. It had another significant increase in brightness in 1892, before weakening again. Since about 1940, its brightness has been steadily increasing. Several models have been proposed to explain what has been happening at η Car [5, 6, 7).

What we propose in this article is to observe Eta Car in another astrophysical window, that of gravitational waves, in order to try to better elucidate what happens in this system. A question then arises: "Can the gravitational waves emitted by Eta Carinae be detected?" In the next section, a way to obtain the detection of the gravitational waves emitted by Eta Car will be proposed.

## 2. Gravitational Waves from Eta Car and the proposed scheme for its detection

The orbital period of the binary existing in η Car is 2,022.7 ± 1.3 days [8], therefore the fundamental frequency of the gravitational waves emitted by this binary is approximately 11.444 nanoHetz ($\log_{10} f$ /Hz = -6.94) which indicates the pulsar timing array (PTA) technique as ideal for its detection. To do this, it is necessary to find the most suitable pulsars to be used in the pulsar timing array technique for detecting gravitational waves from Eta Car. Note that this is not the detection of a background of gravitational waves, but of an isolated but strong source.

η Car is approximately 7500 light years from Earth (2.3 kpc) at celestial coordinates 10h 45m 03.591s and −59º41'04.26'' [9]. Assuming that the masses of the components of the η Car binary have masses around 100 and 30 solar masses, respectively, the amplitude of the gravitational waves from η Car that reach the Earth is approximately $2 \times 10^{-14}$, therefore the residue, calculated by the expression $h/\Omega$, would be around 0.27 microseconds, which would be detectable using the PTA technique.

Table 1. Parameters for some selected pulsars. The four in bold were chosen.

| Pulsar | Period | Pulse Width |
|---|---|---|
| PSR B1257+12 | 6.21853194840048 ms | 660 μs |
| PSR J0348+0432 | 39.1226563571297 ms | 4.6 ms |
| PSR B1937+21 | 1.557806567041489 ms | 10 μs |
| PSR J1748−2446ad | 1.39595482(6) ms | 160 μs |
| PSR J0437−4715 | 5.757451941593412 ms | 4.6 ms |
| PSR J0952−0607 | 1.41379835502312 ms | 212 μs |
| PSR J1719−1438 | 5.7901 ms | 160 μs |
| PSR J2322−2650 | 3.4630991790879 ms | 104 μs |
| PSR J1057−5226 | 197.1 ms | 9.5 ms |
| PSR J1909−3744 | 2.947108069160717 ms | 43 μs |

Table 2. Pulsar distances do Earth and η Car (in light-years). The four in bold were chosen.

| Pulsar | Distance from the pulsar to Earth (l-y) | Distance from the pulsar to η Car (l-y) |
|---|---|---|
| PSR B1257+12 | 2300 | 7311 |
| PSR J0348+0432 | 6800 | 11050 |
| PSR B1937+21 | 11500 | 17242 |
| PSR J1748−2446ad | 18000 | 17781 |
| PSR J0437−4715 | 510 | 7194 |
| PSR J0952−0607 | 4500 | 6109 |
| PSR J1719−1438 | 3900 | 7932 |
| PSR J2322−2650 | 750 | 7578 |
| PSR J1057−5226 | 2350 | 5179 |
| PSR J1909−3744 | 3500 | 7285 |

Table 3. Angle (in degrees) the front of the gravitational wave originating from Eta Car forms with the Earth-pulsar segment, in the vicinity of Earth and in the vicinity of the pulsar. The four in bold were chosen.

| Pulsar | Angle in the vicinity of Earth | Angle in the vicinity of the pulsar |
|---|---|---|
| PSR B1257+12 | 76.43 | 85.8 |
| PSR J0348+0432 | 101.08 | 41.8 |
| PSR B1937+21 | 129.1 | 19.7 |
| PSR J1748−2446ad | 76.27 | 24.2 |
| PSR J0437−4715 | 51.55 | 54.7 |
| PSR J0952−0607 | 54.52 | 88.6 |
| PSR J1719−1438 | 81.6 | 69.3 |
| PSR J2322−2650 | 93.15 | 81.2 |
| PSR J1057−5226 | 7.44 | 10.8 |
| PSR J1909−3744 | 72.92 | 79.7 |

Table 4. Residual-to-pulse-width ratio and time delay (T) of the information in gravitational waves that we receive today on Earth in relation to that received on the pulsar (years)). The four in bold were chosen.

| Pulsar | Residual-to-pulse-width ratio | Time delay (T) |
|---|---|---|
| PSR B1257+12 | 4.15E-04 | 2111 |
| PSR J0348+0432 | 5.95E-05 | 10350 |
| PSR B1937+21 | 2.74E-02 | 21242 |
| PSR J1748−2446ad | 1.71E-3 | 28281 |
| PSR J0437−4715 | 5.95E-05 | 204 |
| PSR J0952−0607 | 1.29E-03 | 3109 |
| PSR J1719−1438 | 1.71E-03 | 4332 |
| PSR J2322−2650 | 2.63E-03 | 828 |
| PSR J1057−5226 | 2.88E-05 | 29 |
| PSR J1909−3744 | 6.37E-03 | 3285 |

In Tables 1, 2, 3, and 4, the parameters of some selected pulsars and other information related to those pulsars are presented. The pulsar periods, pulse widths, and distances to Earth were found in the literature. All other parameters were calculated from these.

Due to the location of Eta Car and the value of the residue, of the more than 60 pulsars commonly used for the PTA technique [10], we suggest the pulsars PSR J0952−0607, PSR J1719−1438, PSR J2322−2650 and PSR J1909−3744 to be part of the search network, because each of them has a relatively high residue/pulse width ratio and a position in space that forms with the Earth and η Car an approximately isosceles triangle (all four pulsars are approximately the same distance from η Car as the Earth is from it) and where at least one of the base (the pulsar-Earth segment) angles is close to 90 degrees. In this way, the front of the gravitational wave originating from η Car forms an angle close to 90 degrees with the Earth-pulsar segment, at least in the vicinity of one of them (Earth or pulsar), optimizing a detection of tensor polarizations.

## 3. Discussion

It is important to emphasize that the pulses received will report the gravitational waves that arrive at Earth today from η Car (which coincides with what is observed in the electromagnetic window today) together with the gravitational waves that arrived at Eta Car T years ago, where T = c x (η Car-pulsar distance + pulsar-Earth distance – η Car-Earth distance) and c the speed of light. For the case of the four proposed pulsars, T is in the range of 828 to 4332 years, as shown in Table 1. The signal will be a composition of the frequencies and phases at these two moments and the polarizations and angles that reach the Earth and the pulsar also at these two moments. The difference between these phases can decrease (or increase) the value of the measured residue. In addition to the fact that Eta Car could have been emitting a wave with different amplitudes and frequencies in the past. It could even be a triple system in the not-too-distant past. In fact, these uncertainties could make this hypothetical detection even more interesting.

It doesn't hurt to try, specially because the data is probably already recorded. The high eccentricity (0.9) of the Eta Carinae binary orbit can help identify the signal in the data due to the presence of emitted gravitational wave harmonics [11].

**Acknowledgments:** The author would like to thank the Brazilian Ministry of Science, Technology and Innovation, and the Brazilian Space Agency (AEB), which supported the present work under PO 20VB.0009. Support from the Conselho Nacional de Desenvolvimento Cientif.co e Tecnol.gico (CNPq) is also acknowledged under Grant Number 301516/2026-0.

## References

1. Damineli, A. The 5.52 year cycle of Eta Carinae. Use code with caution. Astrophys. J. Lett. 1996, 460, L49–L52.
2. Iping, R.C.; Sonneborn, G.; Gull, T.R.; Ivarsson, S.; Nielsen, K. Searching for radial velocity variations in Eta Carinae. In Proceedings of the American Astronomical Society Meeting 207, 2005; Volume 207, p. 1445.
3. Damineli, A.; Hillier, D.J.; Corcoran, M.F.; Stahl, O.; Levenhagen, R.S.; Leister, N.V.; et al. The periodicity of the η Carinae events. Mon. Not. R. Astron. Soc. 2008, 384, 1649–1658.
4. Gull, T.R.; Damineli, A. JD13 – Eta Carinae in the context of the most massive stars. Proc. Int. Astron. Union 2010, 5, 373–398.
5. Damineli, A.; Kaufer, A.;Wolf, B.; Stahl, O.; Lopes, D.F.; de Ara.jo, F.X. η Carinae: Binarity confirmed. Astrophys. J. 2000, 528, L101–L104.
6. McKinnon, D.; Gull, T.R.; Madura, T. Eta Carinae: An astrophysical laboratory to study conditions during the transition between a pseudo-supernova and a supernova. In Proceedings of the American Astronomical Society Meeting 223, 2014; Volume 223, p. 405.03.
7. Smith, N. A model for the 19th century eruption of Eta Carinae: CSM interaction like a scaled-down Type II Supernova. Mon. Not. R. Astron. Soc. 2013, 429, 2366–2379.
8. Damineli, A.; Hillier, D.J.; Corcoran, M.F.; Stahl, O.; Levenhagen, R.S.; Leister, N.V.; et al. The periodicity of the η Carinae events. Mon. Not. R. Astron. Soc. 2008, 384, 1649–1658.
9. H.g, E.; Fabricius, C.; Makarov, V.V.; Urban, S.; Corbin, T.;Wycoff, G.; et al. The Tycho-2 catalogue of the 2.5 million brightest stars. Astron. Astrophys. 2000, 355, L27–L30.
10. Agazie, G.; et al. The NANOGrav 15 yr Data Set: Evidence for a Gravitational-wave Background. Astrophys. J. Lett. 2023, 951, L8.
11. Madura, T.I.; Gull, T.R.; Owocki, S.P.; Groh, J.H.; Okazaki, A.T.; Russell, C.M.P. Constraining the absolute orientation of η Carinae's binary orbit: A 3D dynamical model for the broad [Fe III] emission. Mon. Not. R. Astron. Soc. 2012, 420, 2064–2079.